\documentstyle[sprocl,epsfig]{article}

\def\Journal#1#2#3#4{{#1} {\bf #2}, #3 (#4)}

\def\NPO{{\em Nucl. Phys.}}
\def\PRL{\em Phys. Rev. Lett.}

\def\ra{\rightarrow}
\def\EE{\mathrm{e^+e^-}}
\def\MM{\mu^+\mu^-}

\def\X{\mathrm{X}}
\def\Z{\mathrm{Z}}
\def\bigH{\mathrm{H}}
\def\MH{m_{\mathrm{H}}}
\def\EEZH{\EE\ra\Z\bigH}
\def\ZEE{\Z\ra\EE}
\def\ZMM{\Z\ra\MM}
\def\LL{\mathrm{\ell^+\ell^-}}
\def\ZLL{\Z\ra\LL}
\def\EEGGFF{\EE\ra\EE\mathrm{f\,\bar{f}}}
\def\EFFG{\EE\ra\mathrm{f\,\bar{f}}(\gamma)}
\def\EEWW{\EE\ra\mathrm{W^+W^-}}
\def\EEZZ{\EE\ra\Z\Z}

\def\MeV{\ifmmode {\mathrm{Me\kern -0.1em V}}\else
                   \textrm{Me\kern -0.1em V}\fi}%
\def\GeV{\ifmmode {\mathrm{Ge\kern -0.1em V}}\else
                   \textrm{Ge\kern -0.1em V}\fi}%
\def\TeV{\ifmmode {\mathrm{Te\kern -0.1em V}}\else
                   \textrm{Te\kern -0.1em V}\fi}%

\begin{document}

\title{Measurement of the Higgs Cross Section and Mass with Linear Colliders}

\author{P. Garc\'{\i}a--Abia}

\address{Institut f\"ur Physik der Universit\"at Basel, Klingelbergstrasse 82,\\
         CH--4056--Basel, Switzerland}

\author{W. Lohmann}

\address{DESY, Platanenallee 6, D-15738 Zeuthen, Germany}


\maketitle

\abstracts{We  report on the  accuracy of the  measurement  of the Higgs
           boson mass and the total cross section of the process $\EEZH$
           that would be achieved in a linear  collider  operating  at a
           centre--of--mass energy of 350~\GeV{}, assuming an integrated
           luminosity of  500~fb$^{-1}$.  For that we have exploited the
           recoil mass off the Z using its leptonic decays into electron
           and muon pairs.  The Higgs mass is determined with 150~\MeV{}
           accuracy, the recoil mass resolution is about  1.5~\GeV{} and
           the cross  section is obtained  with a  statistical  error of
           3\%.}

\section{Introduction}

An $\EE$  collider  with  centre--of--mass  energy  $\sqrt{s}$  of a few
hundred~GeV is  particularly  well suited to discover the Higgs boson if
its  mass  is of  $\cal{O}$(100~\GeV{}).  Even  in the  case a  particle
interpreted  as the  Higgs  boson  was  discovered  already  at  another
accelerator like LHC a linear collider is the only facility which allows
to  identify  it  as  the  Higgs  boson   predicted   by  the   Standard
Model~\cite{sm}.  Crucial  tests here are the  values the  couplings  to
gauge bosons and fermions.  A quantity which depends on the couplings to
the gauge  bosons is the Higgs  boson  production  cross  section in the
Higgs-strahlung  or fusion process.  The couplings to different  fermion
species are accessible by the  measurement of the Higgs boson  branching
fractions.  For this the  production  cross  section of Higgs  bosons is
needed.
 
In this letter, we present a method to measure the cross section for the
production  of the Higgs boson  independent  of its branching  fractions
exploiting  the  recoil-mass  method.  We take into account the detector
resolutions, the efficiencies of lepton  identification and the relevant
background processes.

\section{Experimental Conditions}

The  study  is   performed   for  a  linear   collider   operated  at  a
centre--of--mass  energy of  350~\GeV{}.  The  assumed  data  statistics
corresponds to an integrated  luminosity of 500~fb$^{-1}$.  The detector
used in the  simulation  follows  the  proposal  presented  in the TESTA
Conceptual   Design   Report~\cite{cdr}.  

The interaction region is surrounded by a central tracker  consisting of
a  silicon  microvertex  detector  as  the  innermost  part  and a  time
projection  chamber.  In the radial direction follows an electromagnetic
calorimeter, a hadron calorimeter, the coils of a superconducting magnet
and an  instrumented  iron flux return  yoke.  The  solenoidal  magnetic
field is 3 Tesla.  The central tracker momentum resolution is
$\sigma_{p_t}/{p_t} = 7 \cdot 10^{-5} \cdot p_t~(\GeV{})$
and the energy resolution of the electromagnetic calorimeter
$\sigma_{E}/E = 10\%/\sqrt{E} + 0.6 \%$, $E$ in~\GeV.
The polar  angular  coverage  of the  central  tracker  maintaining  the
resolution  is  $|\cos\theta|  < 0.85$,  above  this range the  tracking
resolution  degrades.  The electromagnetic and hadron calorimeters cover
the region  $|\cos\theta| < 0.996$, maintaining  the resolution over the
whole angular  range.  The  simulation  of the detector is done with the
program SIMDET~\cite{simdet}.

\section{Method of the Measurement}

In the energy  range  considered  the  dominant  process for Higgs boson
production  in the  Standard  Model is $\EEZH$.  To determine  the cross
section for this process,  $\sigma(\Z\bigH)$, we exploit leptonic decays
of the Z boson,  $\ZEE$ and $\ZMM$.  These final states  exhibit a clean
signature  in  the  detector.  Hence  they  are  easily   selected  with
selection  efficiencies  expected to be independent of the decay mode of
the Higgs boson.  Due to the  excellent  momentum and energy  resolution
the Z is well reconstructed and the recoil mass against the Z,
$m_{\X}^2  =  s - 2 \cdot \sqrt{s}\cdot E_{\Z} + m_{\Z}^2$,
is  exploited  to detect the Higgs  boson and to measure its  production
cross  section.  Here  $E_{\Z}$ and $m_{\Z}$ are the energy and the mass
of the Z.  Bremsstrahlung and beamstrahlung are taken into account.

\subsection{Signal and Background}

Events  of  the  signal,   $\EEZH$,  are  generated   using  the  PYTHIA
program~\cite{pythia} for Higgs boson masses of 120, 140 and 160~\GeV{}.
The  Standard  Model  cross  sections  and the  expected  event  numbers
corresponding   to  a   luminosity   of   500~fb$^{-1}$   are  given  in
Table~\ref{table:hevents}.

\begin{table}[hb]
\vspace{-0.4cm}
\begin{minipage}[t]{0.47\textwidth}
\begin{tabular}{ccc}
  $\MH$ (\GeV)     &$\sigma$ (fb)& events               \\ \hline
  120              & 5.3         & $2.6\cdot 10^3$      \\
  140              & 4.3         & $2.1\cdot 10^3$      \\
  160              & 3.6         & $1.8\cdot 10^3$      \\ \hline
\end{tabular}
\caption{The  cross  section  for  the  process  $\EEZH\ra\LL\bigH$  for
         different  Higgs boson masses and the number of expected events
         for a luminosity of 500~fb$^{-1}$.}
\label{table:hevents}
\end{minipage}
\hfill
\begin{minipage}[t]{0.47\textwidth}
\begin{center}
\begin{tabular}{lc}
background process    & events               \\ \hline
 $\EEGGFF$            & $2.0\cdot 10^9$      \\
 $\EFFG$              & $2.0\cdot 10^7$      \\
 $\EEWW$              & $7.0\cdot 10^6$      \\
 $\EEZZ$              & $5.0\cdot 10^5$      \\  \hline
\end{tabular}
\caption{\label{table:bevents} The number of events expected for several
         background sources.}
\end{center}
\end{minipage}
\vspace{-0.1cm}
\end{table}

The following  background processes are considered:  $\EEGGFF$, $\EFFG$,
$\EEWW$  and  $\EEZZ$.  The number of events  expected  in each of these
processes is given in Table~\ref{table:bevents}.

\begin{figure}[bth]
\begin{minipage}[t]{0.47\textwidth}
\centerline{\epsfig{figure=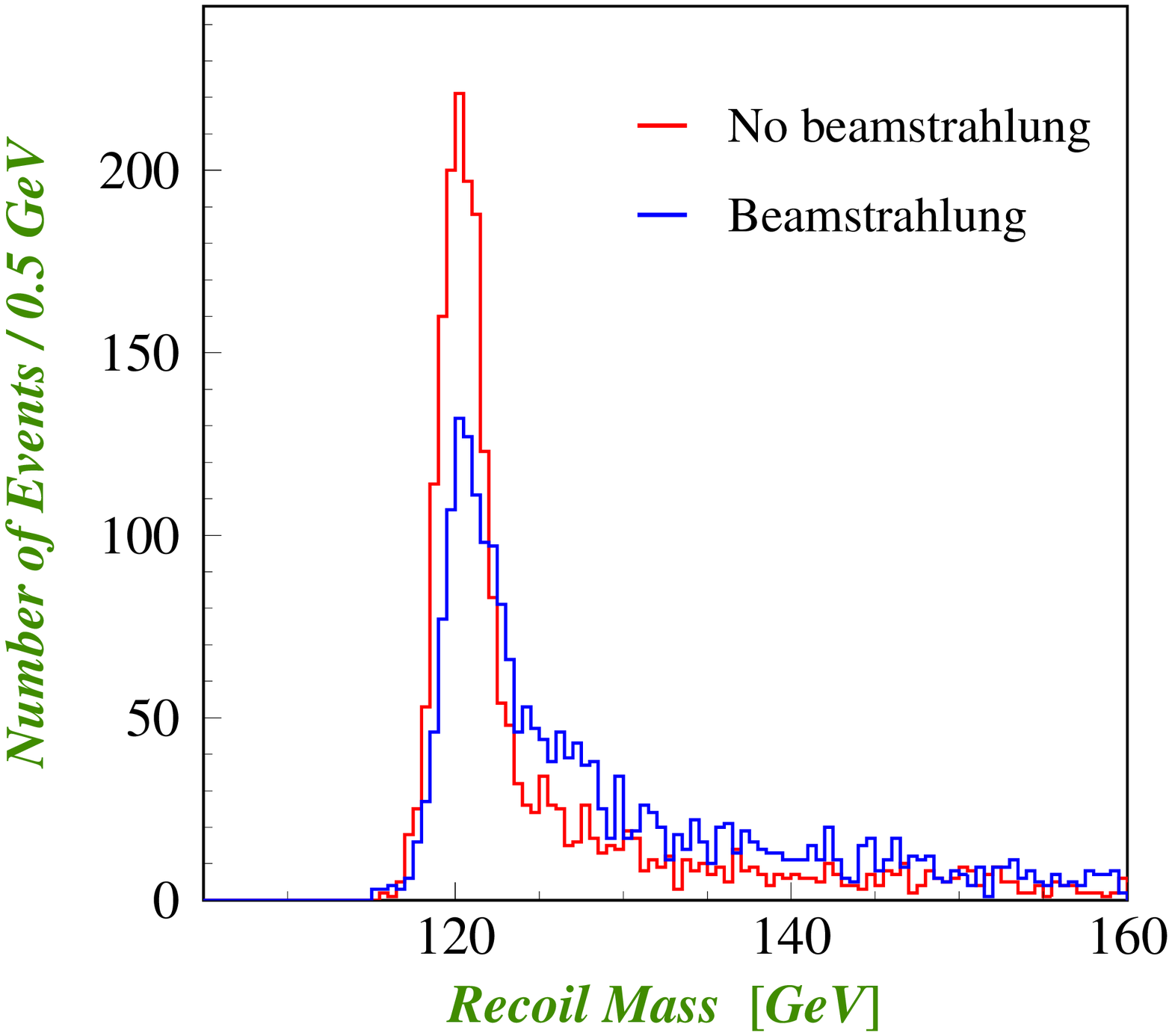,width=1.0\textwidth}}
\caption{\label{beamst} The recoil mass spectra off the Z with and without
         taking into account beamstrahlung.}
\end{minipage}
\hfill
\begin{minipage}[t]{0.47\textwidth}
\centerline{\epsfig{figure=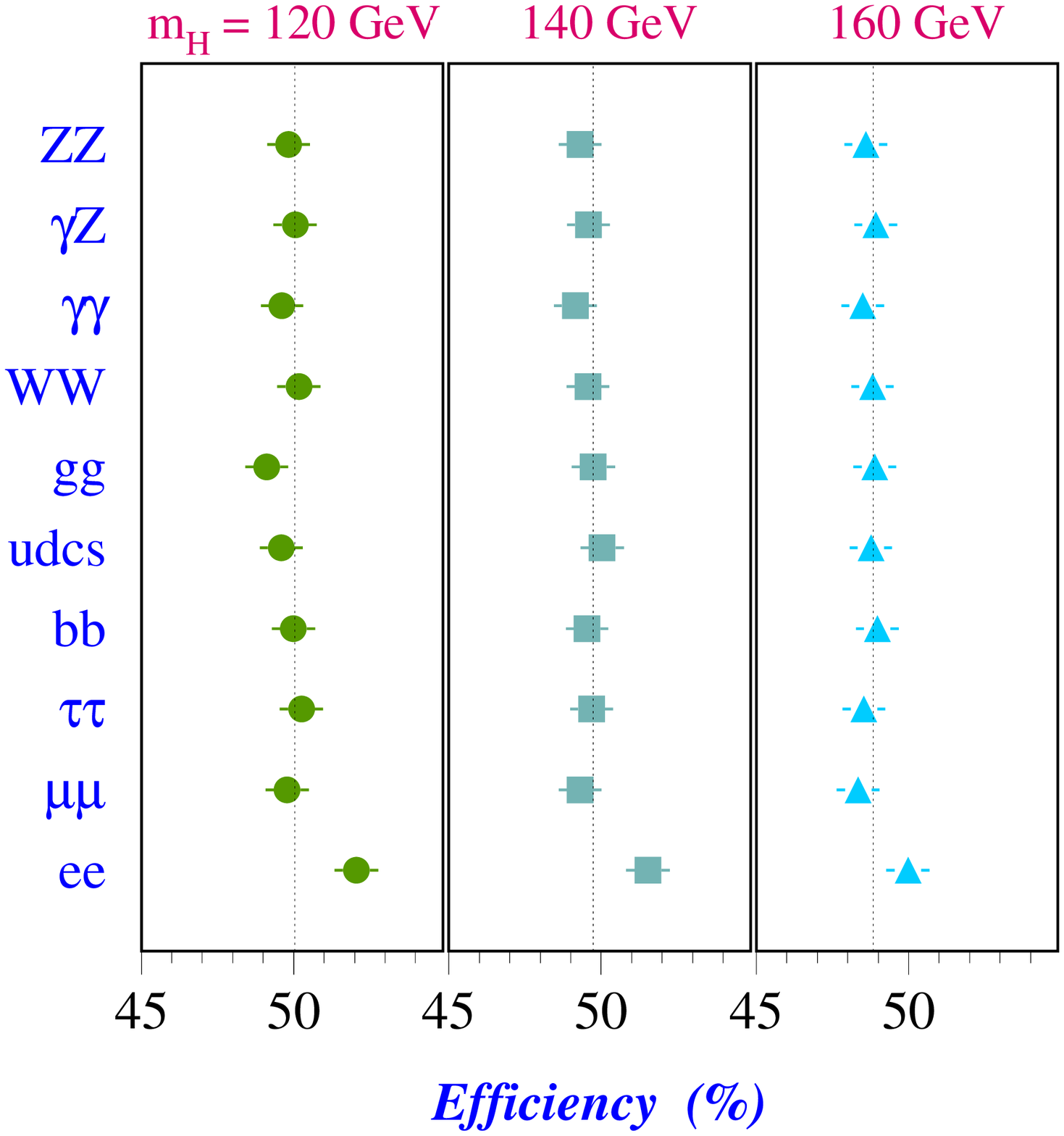,width=1.0\textwidth}}
\caption{\label{cheff}  The selection efficiency of  $\EEZH\ra\EE\rm{X}$
         for several Higgs boson decay modes.}
\end{minipage}
\vspace{-0.3cm}
\end{figure}

Initial state bremsstrahlung is simulated with PYTHIA.  Beamstrahlung is
taken into account using the CIRCE  program~\cite{circe}.  The impact of
beamstrahlung on the recoil mass is illustrated in  Figure~\ref{beamst}.
Beamstrahlung  broadens the recoil boson mass  distribution and hence is
of importance for this study.

Both  signal  and  background  events  are  processed  by  the  detector
simulation  package SIMDET.  The output in terms of reconstructed  track
momenta and  calorimetric  cluster  energies is used for the forthcoming
analysis.

\subsection{Lepton Identification and Selection of $\ZLL$}

Electrons  are  identified  as energy  deposits  in the  electromagnetic
calorimeter  whose  shape  is  compatible  with the  expectation  for an
electromagnetic  shower and with a matched track in the central tracker.
The  measured  track  momentum  and shower  energy must be in  agreement
within 5\% and the shower  leakage into the hadron  calorimeter  must be
less than 2~\GeV.  Muons are tracks  pointing to energy  deposits in the
calorimeters  which are consistent  with the  expectation  for a minimum
ionising  particle.  Both  electrons and muons must have momenta  larger
than 10~\GeV{} and fulfil the polar angle cut $|\cos\theta| < 0.9$.  The
mass of the  leptons is  required to be within  5~\GeV{}  equal to the Z
mass.  To suppress  background from $\EEZZ$, the production  polar angle
of the two--lepton system must be $|\cos\theta_{\ell\ell}| < 0.6$.

The  selection  efficiencies  for  the  processes   $\EEZH\ra\EE\X$  and
$\EEZH\ra\MM\X$   are   listed   in    Table~\ref{table:seleff}.   These
efficiencies  are  independent  on the Higgs boson  decay mode.  This is
demonstrated in Figure~\ref{cheff} for the channel $\EEZH\ra\EE{\rm X}$.
The only  exception  is the case when the Higgs  boson  decays  like the
detected  Z,  $\bigH\ra\EE$.  This mode is  negligible  in the  Standard
Model.  The results for the channel $\EEZH\ra\MM\X$ are very similar.

\begin{table*}[hb]
\vspace{-0.2cm}
\begin{center}
\begin{tabular}{ccc}
 $\MH$ (\GeV)     & $\bigH \EE$ efficiency & $\bigH \MM$ efficiency \\ \hline
   120            &   50.4 $\pm$ 1.0   & 52.8 $\pm$ 1.0           \\
   140            &   49.4 $\pm$ 1.0   & 51.7 $\pm$ 1.0           \\
   160            &   48.8  $\pm$ 1.0  & 51.3 $\pm$ 1.0           \\ \hline
\end{tabular}
\caption{\label{table:seleff}    The    selection    efficiencies    for
         $\EEZH\ra\LL\X$ processes for several Higgs boson masses.}
\end{center}
\vspace{-0.4cm}
\end{table*}

\subsection{Results}

\begin{figure}[bth]
\vspace{-0.1cm}
\begin{minipage}[t]{0.47\textwidth}
\centerline{\epsfig{figure=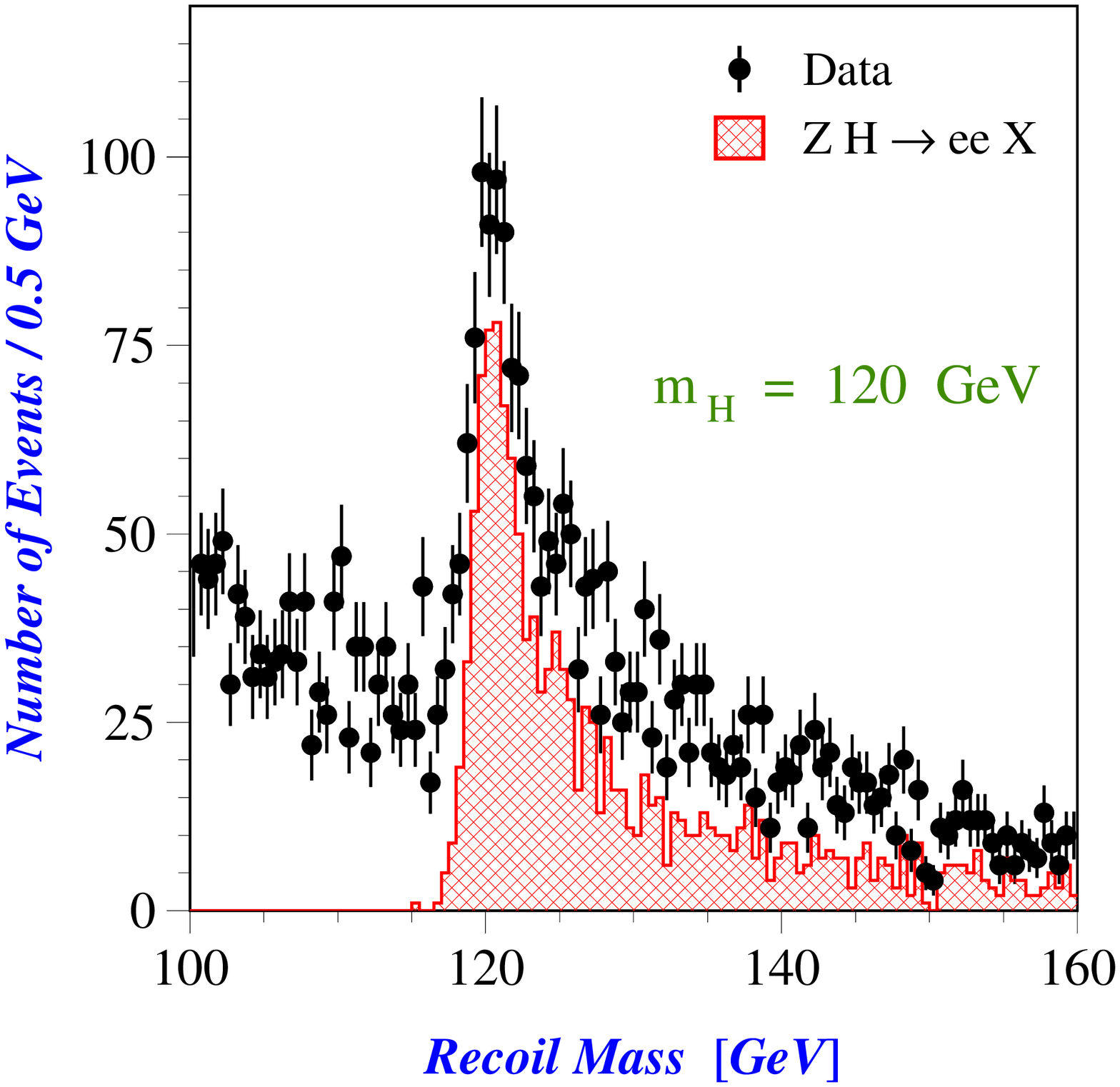,width=1.0\textwidth}}
\vspace{-0.2cm}
\caption{\label{recom}   The   recoil   mass   spectra   off  the  Z  in
         $\EEZH\ra\EE\X$ events.}
\end{minipage}
\hfill
\begin{minipage}[t]{0.47\textwidth}
\centerline{\epsfig{figure=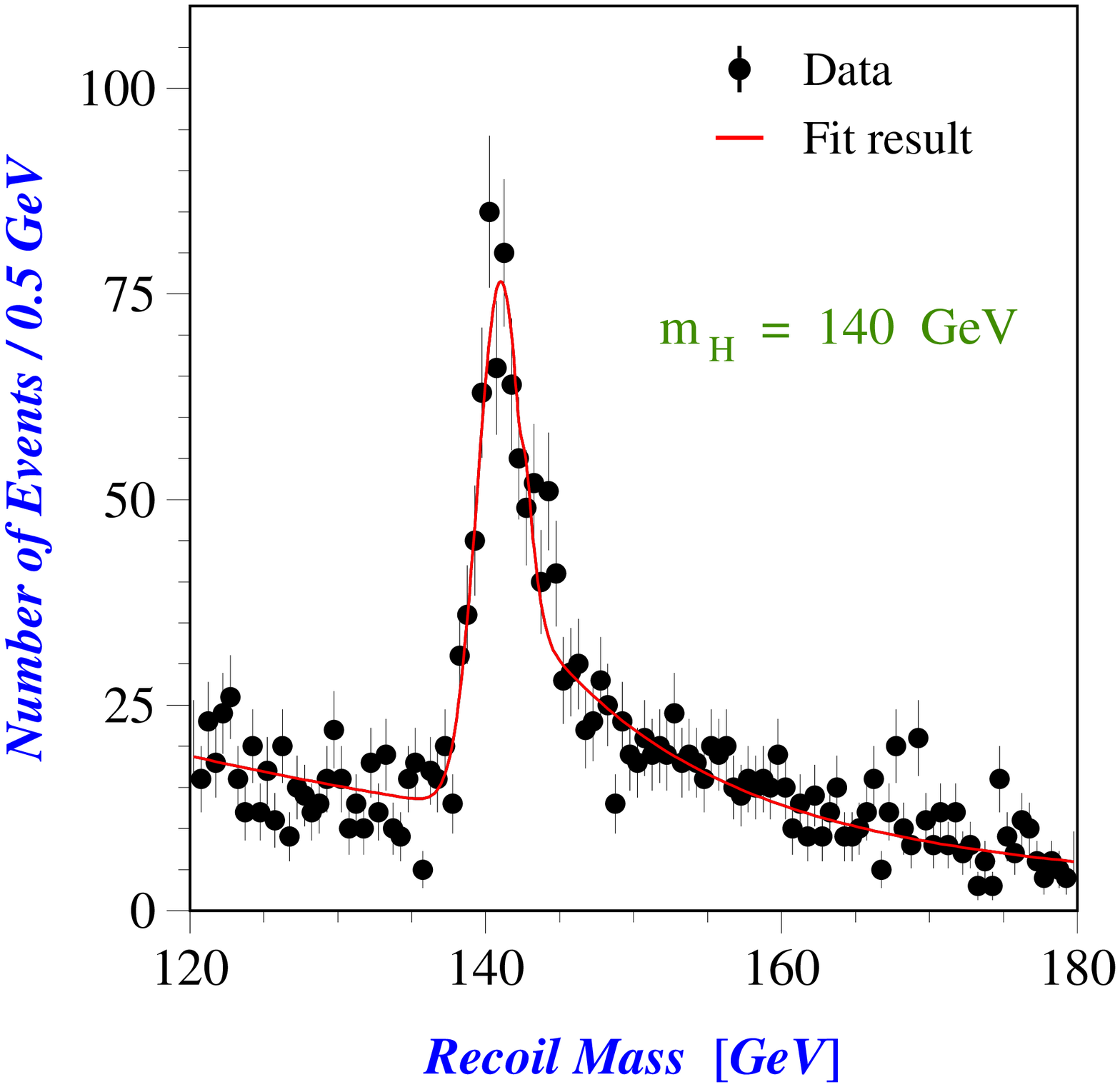,width=1.0\textwidth}}
\vspace{-0.2cm}
\caption{\label{fitreco}  The fit of the recoil mass spectra off the Z
         in $\EEZH\ra\MM\X$ events.}
\end{minipage}
\end{figure}

Figure~\ref{recom}  shows the recoil mass  distribution  obtained  for a
$\MH =$  120~\GeV{}  signal  and the  background  contribution  for  the
channel $\EEZH\ra\EE\X$.  The Higgs boson appears well on top of a small
background, mainly from $\EEZZ$ events.

The recoil mass spectra are fitted with a superposition  of a signal and
a background  distribution.  To account for the asymmetric  signal shape
two approaches are used.  One describes the left side of the signal as a
Gaussian and the right side as the sum of a Gaussian and an  exponential
function.  The  other  uses  an  Edgeworth  expansion~\cite{edge}  of  a
Gaussian.  The background shape is parametrised as an exponential.

The fit is performed with the mass ($m_{\X}$), the width ($\sigma_{\X}$)
and the  normalisation  of the  signal  as free  parameters.  A  typical
result is shown in  Figure~\ref{fitreco}  for a sample  generated with a
Higgs boson mass $\MH =$ 140~\GeV.

The  masses  $m_{\X}$  agree  well with the  values  used in the  signal
generation.  The accuracy of the mass  determination  is about 150~\MeV.
The width of the distribution, reflecting the recoil mass resolution, is
about  1.5~\GeV.  The cross  sections  are obtained  with a  statistical
error   of  3\%  for  the   two   final   states   $\EEZH\ra\EE\X$   and
$\EEZH\ra\MM\X$  combined.  The obtained values for the production cross
sections  are given in  Table~\ref{table:hxsec}.  The  systematic  error
includes  the  uncertainties  from  the  selection  efficiencies  and  a
luminosity measurement error of one 1\%.

\begin{table*}[hbt]
\begin{center}
\begin{tabular}{ccc}
  $\MH$ (\GeV)  & $\sigma$ (fb) $\ZEE$   & $\sigma$ (fb) $\ZMM$       \\ \hline
  120       & 5.26 $\pm$ 0.18 $\pm$ 0.13 & 5.35 $\pm$ 0.21 $\pm$ 0.13 \\
  140       & 4.38 $\pm$ 0.18 $\pm$ 0.11 & 4.39 $\pm$ 0.17 $\pm$ 0.10 \\
  160       & 3.68 $\pm$ 0.17 $\pm$ 0.09 & 3.52 $\pm$ 0.15 $\pm$ 0.08 \\ \hline
\end{tabular}
\caption{\label{table:hxsec}   The  fit  results  for  the  Higgs  boson
         production cross section.}
\end{center}
\vspace{-0.3cm}
\end{table*}

\end{document}